\documentclass[floats,twocolumn,prl,aps]{revtex4}

\usepackage{graphicx}

\newcommand\beq{\begin{equation}}
\newcommand\eeq{\end{equation}}
\newcommand\bea{\begin{eqnarray}}
\newcommand\eea{\end{eqnarray}}
\newcommand{\nonum}{\nonumber}

\begin{document}

\title{\bf Quantum spin pumping  
at fractionally quantized magnetization
state for a system with competing exchange interactions.}

\author{\bf Sujit Sarkar$^{1,2}$ ~and ~\bf C. D. Hu$^{2}$}
\affiliation {\it 1. PoornaPrajna Institute of Scientific Research,
4 Sadashivanagar, Bangalore 5600 80, India\\
2. Department of Physics,     
The National Taiwan University, Taipei 10617, Taiwan, R. O. C}

\date{\today}

\begin{abstract}
We study the quantum spin pumping of an antiferromagnetic spin-1/2 chain with
competing exchange interactions. 
We
show that spatially periodic potential modulated in space and time
acts as a quantum spin pump. In our model system, an applied electric
field causes a spin gap to its critical ground state by introducing
bond-alternation exchange interactions.
We study quantum spin pumping at different quantized magnetization states
and also explain physically the presence and absence of quantum spin pumping
at different fractionally quantized magnetization states.\\
\end{abstract}

\maketitle


\section{ 1. Introduction}
An adiabatic quantum pump is a device that generates a dc current by a 
cyclic variation of some system parameters, the variation being slow
enough so that the system remains close to the ground state throughout
the pumping cycle. The pumping physics gets more attraction after the
pioneering work of Thouless \cite{thou1,thou2}. 
Quantum adiabatic pumping physics is not only related to the
spin system but also related to the other systems like open quantum dots 
\cite{brou,alt,dot},
superconducting quantum wires \cite{bla,wire}, 
the Luttinger quantum wire \cite{lutt1} and also to the interacting
quantum wire \cite{lutt2}.\\
The motivation of our study is the following: we have understood from the previous
paragraph that adiabatic quantum pumping may arise in different
systems due to the presence of different pumping sources. Here we would like to
study the adiabatic quantum pumping of a system that 
has not covered in any one of the 
previous studies.
We consider an antiferromagnetic spin-1/2 chain with competing exchange
interactions to study the adiabatic quantum spin pumping.
We consider both nearest-neighbor (NN) and next-nearest-neighbor (NNN) 
exchange interaction in the 
spin chain and only consider the
presence of electric field that induces time dependent dimerization in 
both the exchange interaction. We also study the spin pumping at different states
of magnetization. Our approach is completely analytical. We use Abelian bosonization
and one-loop RG calculation to explore spin pumping physics of this model
Hamiltonians system. Shindou \cite{shin} has studied only Heisenberg XXZ spin chain
with NN dimerization. There
are no competing exchange interactions  
and also the effect of different
states of magnetization on adiabatic spin pumping physics is absent.
Shindou \cite{shin} has considered 
two perturbations which opens a gap in the excitation spectrum. One of them 
is the bond-alternation exchange interaction which leads to the dimerized
state and the other one is the staggered magnetic field which locks the spin 
into a Neel ordered state.  
In his model, applied cyclic
electric and magnetic field control staggered component of exchange interaction
and staggered magnetic field respectively. In our case there is only one 
perturbation which opens gap in the excitation spectrum. 
A part of our model has some experimental relevance \cite{koh,ful}
. Suppose we have a spin-1/2
chain (like Cu-benzoate and the charge order phase of Y$b_4$A$s_3$) with unit
cell containing two crystallographic inequivalent sites, where both the translation
symmetry ( T $ S_j \rightarrow S_{j+1}$ ) and the bond centered inversion symmetry
( $I_{bond}$ ${S}_{i-j} \leftrightarrow S_{i+j+1}$ ) are crystallographically
broken. But the NN exchange interaction $J$ in these spin chain does not
have any alternating component because the system has site-centered inversion
symmetry which exchanges the NN bonds. If we apply electric field in a particular
direction of the system \cite{shin2}, then 
we may break the site-centered inversion symmetry and
it yields the bond alternation component in the NN exchange interaction. The
additional interactions like, NNN exchange interaction and it's 
alternating components in the
Hamiltonians are completely theoretical. We consider these terms in 
the Hamiltonians to study the nontrivial and interesting effects of 
these terms over the basic interactions.  
In sec II. we present the model Hamiltonians and general
derivations. Different subsections are for the different states of magnetization.
Sec.III is devoted for conclusions.

\section{ 2. Model Hamiltonians and Continuum Field Theoretical Study:}

In our model Hamiltonians, we consider the presence of time dependent
bond-alternation (dimerization) in both NN and NNN exchange interactions. 
We assume that the 
time dependence
of dimerization is restored by the applied alternating electric field. 
In this section we do the all calculations, different subsections
are for the special limit of this general derivations. 
Our model Hamiltonians
are the following.   
\bea
H_A &=& J_1 \sum_{n}~( {{S}_n}(x)  {{S}_{n+1}(x)}~+~
{{S}_n}(y)  {{S}_{n+1}(y)} ~, \nonum \\
&&+ \Delta {{S}_n}(z)  {{S}_{n+1}(z)} )+ 
 J_2 \sum_{n} (1 - \delta (t) (-1)^n) \vec{S}_n \cdot \vec{S}_{n+2} \nonum \\
&& -{g {\mu}_B H} \sum_{n} {S_n} (z)
\label{hamB}
\eea 
\bea
H_B &=& J_1 \sum_{n} (1 - \delta (t) (-1)^n)~( {{S}_n}(x) {{S}_{n+1}(x)}~+~
{{S}_n}(y) {{S}_{n+1}(y)} , \nonum \\
& & +\Delta~~ {{S}_n}(z) {{S}_{n+1}(z)}~~) + 
J_2 \sum_{n} \vec{S}_n \cdot \vec{S}_{n+2} \nonum \\
& & -{g {\mu}_B H} \sum_{n} {S_n} (z)
\label{hamA}
\eea

where n is the site index, x, y, and z are components of spin. 
$J_1$ and $J_2$ are the nearest-neighbor and next-nearest-neighbor
exchange coupling between spins, ${J_1}, {J_2}~\geq 0$, 
$\Delta$ is z component anisotropy of NN exchange interaction, 
$\delta (t)$ is dimerization
strength, which appears as a time dependent parameter in our Hamiltonians,  
$H$ is the externally applied static 
magnetic field in the z direction. 
The staggered component of exchange interaction is arising due to the
broken site centered inversion symmetry under a electric field in
a particular direction \cite{shin}. A site-centered inversation operation
with the sign of elecrtic field reversed that requires, $\delta$ must be 
an odd function
of electric field \cite{shin}.\\
One can express
spin chain systems to a spinless fermion systems through 
the application of Jordan-Wigner transformation. In Jordan-Wigner transformation
the relation between the spin and the electron creation and annihilation operators
are 
$ S_n^z  =  \psi_n^{\dagger} \psi_n - 1/2 ~$, 
$ S_n^-  =   \psi_n ~\exp [i \pi \sum_{j=-\infty}^{n-1} n_j]~$, 
$ S_n^+  =  \psi_n^{\dagger} ~\exp [-i \pi \sum_{j=-\infty}^{n-1} n_j]~$,
where $n_j = \psi_j^{\dagger} \psi_j$ is the fermion number at site $j$. 
\bea
{H}_{A1}~&=& - \frac{J_1}{2} ~\sum_n ~(\psi_{n+1}^{\dagger} \psi_n + 
\psi_n^{\dagger} \psi_{n+1}) \nonum \\
&& + J_1 \Delta \sum_n (\psi_n^{\dagger} \psi_n - 1/2) (\psi_{n+1}^{\dagger} 
\psi_{n+1} - 1/2) ~,\nonum \\
&&  -g {{\mu}_B} H \sum_{n} (\psi_n^{\dagger} \psi_n - 1/2).
\label{ha1}
\eea
\bea
H_{A2}~&=& J_2 ~\sum_n ~( \psi_{n+2}^{\dagger} \psi_n + {\rm h.c.})
(\psi_{n+1}^{\dagger} \psi_{n+1} - 1/2) \nonum \\
&& +~ J_2 ~\sum_n ~(\psi_n^{\dagger} \psi_n - 1/2)
(\psi_{n+2}^{\dagger} \psi_{n+2} - 1/2) .
\label{hA3}
\eea
\bea
H_{A3}~&=& -~J_2 \delta (t) \sum_n (-1)^n ( \psi_{n+2}^{\dagger} \psi_n + {\rm h.c.})
(\psi_{n+1}^{\dagger} \psi_{n+1} - 1/2) \nonum \\
&& - ~ J_2 \delta (t) \sum_n (-1)^n ~(\psi_n^{\dagger} \psi_n - 1/2)
(\psi_{n+2}^{\dagger} \psi_{n+2} - 1/2).
\label{hb3}
\eea
There is a difference between the first term of Eq.3 with the first
term of Eq.4 and Eq.5.  
This difference arises due to the presence
of an extra factor $e^{- i \pi {n}_{j+1}}$ in the string of
Jordan-Wigner transformation for NNN exchange interactions.

Similarly one can also recast the spin-chain systems with $J_1$ 
dimerization into the spinless fermions. The Hamiltonians are converted as 
follows:
$H_{B1}~=~H_{A1}$, $H_{B2}~=~H_{A2}$ and 
\bea
{H}_{B3}~&=& \frac{J_1}{2} ~\delta (t) \sum_n (-1)^n ~(\psi_{n+1}^{\dagger} \psi_n + 
\psi_n^{\dagger} \psi_{n+1}) \nonum \\
&& + {J_1}{\Delta} {\delta (t)} \sum_n (-1)^n 
(\psi_{n+1}^{\dagger} 
\psi_{n+1} - 1/2)
(\psi_{n}^{\dagger} 
\psi_{n} - 1/2)
\label{ha2}
\eea
Here our Hamiltonians are different from previously studied dimerization problem.
In Ref. \cite{hal} and Ref. \cite{aff} have studied intrinsic dimerization
for frustrated spin-1/2 antiferromagnetic chain. In these studies, there is
no explicit dimerization. Totsuka \cite{tot} and Tonegawa \cite{tone} have studied
the $H_B$ Hamiltonian only. These is no spin pumping physics in any 
one of the previous
studies \cite{hal,aff,tot,tone}. 
There are few other studies \cite{chen,cap,sar1}
based on model $H_A$, but there is no
spin-pumping physics in any one of these studies.
So the current work is
more wide and advance.
Our approach is completely analytical, i.e., we explain
the basic understanding of spin pumping physics of our
model system.
Before we proceed further for continuum field theoretical 
study of these model Hamiltonians, we would like to explain the
basic aspects of quantum spin pump of our model Hamiltonians: 
An adiabatic sliding motion of one dimensional potential,
in gapped fermi surface (insulating state), pumps an integer numbers
of fermions 
per cycle.
In our case the transport of Jordan-Wigner
fermions (spinless  fermions) is nothing but the transport of spin from one end
of the chain to the other end because the number operator of spinless fermions
is related with the z-component of spin density \cite{cal}. 
We shall see that non-zero ${\delta} (t)$ introduces the gap at 
around the
Fermi point and the system is in the insulating state (Peierls insulator).
In this phase spinless fermions form the bonding orbital between the
neighboring sites, which yields a valance band in the momentum space.
It is well known that the physical behavior of the system is identical
at these two Fermi points. From the seminal paper of Berry \cite{berry}, 
One can analyse this double
degeneracy point. It appears as source and sink vector fields defined 
in the generalized
crystal momentum space \cite{berry}.
${B_n} (K) = {{\nabla}_K} \times {A_n} (K)$, and 
${A_n} (K) = \frac{i}{2 \pi} <n (K)| {{\nabla}_K} | n(K)>$, 
where $K = (k, {\delta} (t) )$.
Here $B_n$ and $A_n$ are the fictitious magnetic field (flux) and 
vector potential of the
nth Bloch band respectively.
The degenerate points behave as a magnetic monopole in the generalized
momentum space, whose magnetic unit can be shown to be 1 \cite{shin}, 
analytically
\beq
\int_{S_1} ~dS \cdot {B_{\pm}}~=~ \pm 1
\eeq 
Positive and negative signs of the above equation are respectively
for the conduction and valance band. Conduction and valance bands 
meet at the degeneracy points. 
$S_1 $ represent an arbitrary closed surface which enclose the 
degeneracy point. 
In the adiabatic process the parameter ${\delta} (t)$ is changed
along a loop ($\Gamma$) enclosing the origin (minima of the system).
It is well known in the literature of adiabatic quantum pumping physics
that two independent parameters are needed to achieve the adiabatic 
quantum pumping
in a system \cite{brou}. Here one may consider these two parameters as the real and
imaginary part of the fourier transform of dimerized potential. 
When the shape of the dimerized
potential will change in time, then it amounts to change the phase and
amplitude in time. The role of adiabatic parameters are not explicit
in our study. We define the expression for spin current ($I$) from the analysis
of Berry phase.   
Then according to the original
idea of quantum adiabatic particle transport \cite{thou1,thou2,shin,avron}, 
the total number of spinless fermions ($I$)
which are transported from one side of this system to the other is equal to the
total flux of the valance band, which penetrates the 2D closed sphere 
($S_2 $) spanned by
the $\Gamma$ and Brillioun zone \cite{shin}. 
\beq
 I = \int_{S_2} dS \cdot B_{+1} ~=1
\eeq
We have already understood that quantized
spinless fermion transport is equivalent to the spin transport \cite{cal}.
We will interpret this equation more physically at the end (Eq. 18) of the
next section. 
This quantization is
topologically protected against the other perturbation as long as the gap along the
loop remains finite \cite{avron,shin}. 
 
In the following paragraph, we do the continuum field theoretical
studies of spin pumping 
for different magnetization states and explain the
stabilization of quantized spin pumping against z-component of exchange interactions,
and also from the intrinsic dimerization (when ${J_2} > 0.241 {J_1}$ \cite{hrk}). 
We recast the spinless
fermions operators in terms of field operators by this relation 
\beq
{\psi}(x)~=~~[e^{i k_F x} ~ {\psi}_{R}(x)~+~e^{-i k_F x} ~ {\psi}_{L}(x)]
\eeq
where ${\psi}_{R} (x)$ and ${\psi}_{L}(x) $ describe 
the second-quantized fields of right- and 
left-moving fermions respectively.  
In absence of magnetic field ($H=0$), $k_F~=~\pm {\pi}/2$,
however we are interested to study the systems in presence of static magnetic field. 
Therefore we keep Fermi 
momentum as arbitrary $k_F$.
One can simply absorb the finite magnetization in a shift of field $\phi$ by
$\phi~=~{\tilde {\phi} - \pi m x}$, where $m~=<S_z>$ .
In presence of
magnetic field Fermi momentum and magnetization ($m$) are related by this relation,
$k_F~=~~\frac{\pi}{2} ( 1~-~2 m)$ \cite{gia2}.
We want to express the fermionic fields in terms of bosonic field by this relation 
\beq
{{\psi}_{r}} (x)~=~~\frac{U_r}{\sqrt{2 \pi \alpha}}~~e^{-i ~(r \phi (x)~-~ \theta (x))} 
\eeq
$r$ is denoting the chirality of the fermionic fields,
 right (1) or left movers (-1).
The operators $U_r$ are operators that commute with the bosonic field. $U_r$ of different species
commute and $U_r$ of the same species anticommute. $\phi$ field corresponds to the 
quantum fluctuations (bosonic) of spin and $\theta$ is the dual field of $\phi$. 
They are
related by this relation $ {\phi}_{R}~=~~ \theta ~-~ \phi$ and  $ {\phi}_{L}~=~~ \theta ~+~ \phi$.

Using the standard machinery of continuum field theory \cite{gia2}, 
we finally get the bosonized Hamiltonians
as 
\bea
H_{0}~&=&~v_0 \int_{o}^{L} \frac{dx}{2 \pi} \{ {\pi}^2 : {\Pi}^2 : ~+~ :[ {\partial}_{x} \phi (x) ]^2 : \nonum\\
&&~+~ \frac{g_1}{{\pi}^2 }\int ~ dx~:[ {\partial}_{x} {{\phi}_L} (x) ]^2  :
+ :[ {\partial}_{x} {{\phi}_R} (x) ]^2 : \nonum\\
&&~+~ \frac{g_2}{{\pi}^2 }\int ~ dx~ 
 ({\partial}_{x} {{\phi}_L} (x) ) ({\partial}_{x} {{\phi}_R} (x) ) 
\label{bos1}
\eea
$H_{0}$ is the gapless Tomonoga-Luttinger liquid part of the Hamiltonian
with $v_0~=sink_F$. The analytical expressions for $g_1$ and $g_2$
(related with the forward scattering of fermionic field) are the following. 
$ g_1~=~2 (\Delta ~-~ 2 J_2) ~{sin}^2 k_F ~+~ 2~J_2 ~ sin 2 k_F (\pi~+~sin 2 k_F)$  
$ g_2~=~4 (\Delta ~-~ 2 J_2) ~{sin}^2 k_F ~+~ 4~J_2 ~ {sin}^{2} {2 k_F }$.   

Analytical expressions for 
different exchange interactions of Hamiltonian, $H_A$, are the following.

\beq
H_{J2C1}~=~ \frac{J_2}{2 {\pi}^2 {\alpha}^2}
\int dx :cos[4 \sqrt{K}\phi (x) ~-~ (G -4 k_F )x~-~ 4 k_F a]: .
\eeq
\beq
H_{J2C2}~=~\frac{J_1 \Delta}{2 {\pi}^2 {\alpha}^2}
\int dx :cos[4 \sqrt{K}\phi (x) ~+~ (G - 4 k_F )x~-~2 k_F a]: .
\eeq
\beq
H_{J2C3}=  \frac{J_2 \delta (t)}{2 {\pi}^2 {\alpha}^2}
 \int dx :cos[(\pi- 4 k_F)x~+~4 \sqrt{K}~\phi (x) ~-~4 k_F a]:.
\eeq
Where $G$ is the reciprocal lattice vector. 
Eq.12 and Eq.13 are presenting the umklapp scattering term
from the NN and NNN antiferromagnetic exchange interaction, Eq.14
is appearing due to the presence of dimerized interaction.
Similarly one can also find the expressions for $H_B$ Hamiltonian.
Analytical expressions for $K$ is the following.
\beq
K = {\Big[}\frac{1~-~(8/\pi)~J_2 ~{sin}^2 k_F ~+~ 4 J_2~cos k_F}
      {1~+~(4/\pi) \Delta sin k_F~+~ 4 J_2 ~ 
cos k_F ( 1~+~2/\pi  sin {2 k_F }) }{\Big]}^{1/2}.
\eeq
$v_0$ and $K$ are the two Luttinger liquid parameters.
During this derivation we have used the following relations: 
${\rho}_{R/L} ~=~\frac{-1}{\pi}{\partial}_{x} {\phi}_{R/L} (x)$ and
$[{\phi} (x) ~,~\Pi (x) ]~= i \delta (x-x')$, where
$\Pi (x)~=\frac{1}{\pi} \nabla \theta (x)$,
is the canonically conjugate momentum.
We have also used the following equations,
$S^{z} (x)~= ~a ~[ ~\rho (x)~+~ (-1)^j ~ M(x) ~] ~$.
The bosonized expressions for $\rho$ and $M$ are given by
$ \rho (x) = ~-~ \frac{1}{\sqrt \pi} ~\partial_x \phi (x) ~$, 
$ M(x) = ~\frac{1}{\pi a} ~\cos (2  \phi (x)) ~$.
Similarly one can calculate the analytical expressions for $J_1$ dimerization.
Here we have expressed our all expressions in terms of bare phase field
($\phi$), by
using the conventional practice of continuum field theory \cite{gia2}. 
During these derivations we assume
that $J_1 \gg J_2 ,~ \delta$. $J_2$ is in the unit of $J_1$. Here we neglect the
higher order
of $a$ than $a^2$.\\

\section{ 2.1  Calculations and Results for m=0 Magnetization States:}
At first we discuss $m=0$ magnetization state, it corresponds $k_F ~=~\pm~ \pi/2$.
Here we study both the effect
of XXZ anisotropy ($\Delta$) and the spin-Peierls dimerization (${\delta} (t)$).
The effective Hamiltonian for $J_2$ dimerization become, 
\beq
H_A = H_0 ~+~( \frac{J_2 ~-~{J_1} \Delta}{2 {\pi}^2 {\alpha}^2} )~\int dx 
:cos[4 \sqrt{K}~\phi (x)]: .
\label{ha0}
\eeq
In this effective Hamiltonian (Eq.16), there is no contribution
from dimerized
interaction ($ {k_F} = \pi/2$ limit of Eq. 14),
due to the oscillatory nature of the integrand
(it
leads to the vanishing contribution.).
But the contribution of dimerized potential is present
in the NN exchange interaction. 
Similarly the effective Hamiltonian for $J_1$ dimerization become, 
\bea
H_B & = & H_0 ~+~ ( \frac{J_2 ~-~ {J_1} \Delta}{2 {\pi}^2 {\alpha}^2} )~
\int dx :cos[4 \sqrt{K}~\phi (x)]: \nonum\\
&& ~+~\frac{{J_1} {\delta} (t)}{2 {\pi}^2 {\alpha}^2}\int dx 
:cos[(2 \sqrt{K}~\phi (x)]: .
\label{hb0}
\eea
This dimerization contribution for NN exchange interaction has
originated from the XY interaction. This dimerization is the
spontaneous dimerization, i.e., infinitesimal amount of
${\delta} (t)$ is sufficient to produce a gap around the
Fermi points. 
The other two contribution of $J1$ dimerization are from XXZ 
anisotropy of NN exchange interaction and z-component of NNN
interaction. 
Fig. 1 shows the variation of $K$ with $\frac{J_2}{J_1}$
for different values of $\Delta$. 
We observe from the figure that $K$ is the function of $\Delta$ for fixed
$\frac{J_2}{J_1}$ (inset of Fig. 1 shows the appearence of intrinsic
dimerization as a function of $\frac{J_2}{J_1}$ and $\Delta$. There are
a few studies \cite{nom,ali} on the phase seperation between the spin-fluid and
dimer order phase of frustrated spin chain, but our Hamiltonians are different
from them). 
The second term of Eq. 17 is irrelevant when $K$ is greater
than 1/2. 
So in this parameter space only the time dependent dimerizing field
(third term of Eq. 17)
is relevant and lock the phase operator at ${\phi} = 0 + \frac{n \pi}{\sqrt{K}}$.
Now the locking potential slides adiabatically (here the cyclic electric
field that produces the dimerization). Speed of the sliding potential is low
enough such that system stays in same valley, i.e., there is no scope to jump
onto the other valley.  
The system will acquire $2 \pi$ phase during one
complete cycle of external electric field around the loop encircling 
the minima of critical ground state,
produced by the dimerizing field. This is the basic mechanism of spin pumping of
our system. 
This 
expection is easily verified when we notice the physical meaning of the phase
operator ($\phi$ (x)). Since the spatial derivative of the phase operator 
corresponds to the z-component of spin density, this phase operator is
nothing but the minus of the spatial polarization of the z-component of
spin, i.e., 
$ P_{s^z}~= - \frac{1}{N} \sum_{j=1}^{N} j {S_j}^z $. Shindou
has shown explicitly 
the equivalence between these two consideration \cite{shin}. During the
adiabatic process $ < {\phi}_{t} >$ changes monotonically and acquires
- $2 \pi$ phase. In this process $ {P_s}^{z} $ increases by 1 per cycle.
We define it analytically as 
\beq
{\delta} {P_s}^{z} = \int_{\Gamma} d {P_s}^{z} 
= - \frac{1}{2 \pi} \int dx {{\partial}_x} <{\phi} (x) > = 1
\eeq
This physics always hold as far as the system is locked by the sliding 
potential and ${\Delta} < 1$ \cite{shin}.
This equation (Eq. 18) for spin transport is physically consistent 
with the Eq. 7 (based on Berry phase analysis) of spin current. 
The quantized spin transport of this scenario can be generalized up to the 
value of $\Delta$  for which 
$K$ is greater than 1/2  
. In this limit, z-component
of exchange interaction and also the intrinsic dimerization 
has no effect on the spin pumping
physics of applied electric field induced dimerized interaction of 
$H_B$ Hamiltonian.  
\begin{figure}
\includegraphics[scale=0.35,angle=0]{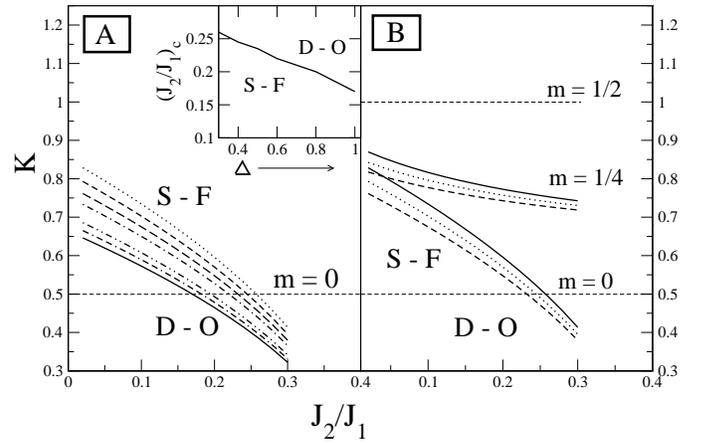} 
\caption{Luttinger liquid parameter ($K$) versus $\frac{J_2}{J_1}$ for different magnetization
state. A constant dashed line at $K=0.5$ is for eye guide line, which separating 
the spin fluid and dimer order instability state. 
D-O: dimer order instability state, S-F: spin fluid Luttinger liquid phase.
{\bf A:} Here we only focus at $m=0$ magnetization state.
Uppermost (dotted) curve is for $\Delta =0.3$ and the lowermost (solid) curve is
for $\Delta =1$. The intermediate curves are for $\Delta = 0.4, 0.5, 0.6.0.8, 0.9$ respectively
from upper to lower one. Inset shows the separation between the spin fluid and
dimer order instability state by a critical line. Here we present the shift of the
${(\frac{J_2}{J_1})}_c$ with $\Delta$. {\bf B:} Here we present the 
curves for different magnetization states. $m=1/2$ is independent of $\Delta$ and 
$\frac{J_2}{J_1}$ in contrast with $m=1/4 , 0$. The curves for $m=1/4 , 0$ plateaus are
for $\Delta = 0.4, 0.5, 0.6$ respectively from upper to lower one.} 
\label{Fig. 2}
\end{figure}

\subsection{ 2.2 Calculations and Results for $m=\frac{1}{4}$ Magnetization States:}

Here we discuss the physics of quantum spin pumping of a finite  
magnetization state. 
We are considering the magnetization state at $ m=\frac{1}{4}$, it corresponds
$ k_F ~=~\pm \frac{\pi}{4}$. 
The effective Hamiltonian for $J_2$ dimerization become,
\bea
H_A & = & H_{0} ~-~ ( \frac{{J_1}\delta (t)}{2 {\pi}^2 {\alpha}^2} )~
\int dx :cos[4 \sqrt{K}~\phi (x)]:
\nonum\\
&& + \frac{A}{2 {\pi}^2 {\alpha}^2}~\int dx :cos[4 \sqrt{K}~\phi (x)~+~\beta]:. 
\label{hj2}
\eea
Where $A~=~\sqrt{ {{{J_1}^2}{\Delta}}^2  ~+~ {J_2}^2 }$, 
${\beta}~=~tan^{-1} \frac{{J_1}{\Delta}}{J_2} $. 
Apparently it appears from the general derivation of section (II), 
that the second and third terms of Eq. 19, will be
absent due to the oscillatory nature of the integrand but this is not the case
when one consider the dimerized lattice. In dimerized lattice, reciprocal
lattice vector $G$ will change from $2 \pi$ to $\pi$ due to the change of the 
size of the 
unit cell. It become more clear, if one write these terms as
$ ~\int dx :cos[ (G-4 k_F)x ~+~ 4 \sqrt{K}~\phi (x)]:$.

Similarly one can write the effective Hamiltonian for $J_1$ dimerization:
\bea
H_B & = & H_{0} ~+~ ( \frac{{J_1} {\Delta} \delta (t)}{2 {\pi}^2 {\alpha}^2} )~
\int dx :sin[4 \sqrt{K}~\phi (x)]:
\nonum\\
&& - \frac{A}{2 {\pi}^2 {\alpha}^2}~\int dx :cos[4 \sqrt{K}~\phi (x)~+~\beta]:. 
\label{hj3}
\eea
The analytical structure of Eq. 19 and Eq. 20 are the same, i.e., the coefficient
of the field ${\phi}(x)$ is the same for all sine-Gordon coupling terms.
The renormalization group equations for these type of interactions 
are \cite{gia2,kos}.
\beq
\frac{dK}{dlnL}~=~ -4 {\pi}^2  K^2 {\delta (t) }^2
\eeq 
\beq
\frac{d \delta (t)}{dlnL}~=~ (2~-~4 K) \delta (t)
\eeq 
It appears from these RG equations that to get a relevant perturbation, 
$K$ should be less than $1/2$.
It reveals from Fig. 1B that $K$ is exceeding the relevant 
value in our region of interest
to mature criteria for spin pumping. 
So the dimerization strength should exceed some critical
value (${\delta}_c$) to initiate the spin pumping phase.  
These two equations are the Kosterlitz-Thousless equation 
for the system in this limit. At the critical point,
system undergoes Kosterlitz-Thouless transition \cite{gia2,kos}.
Since the system flows to the strong coupling (dimer-order) as the 
dimerization strength exceeds some critical
value (${\delta}_c$) initially, we have to guess the physics of this phase. 
We analyze the system in the
limit $\delta~ \rightarrow \pm \infty$ and $K~ \rightarrow 0$. 
In this limit all sine-Gordon couplings are relevant but 
the value of $\phi$ is pinned at the minima of $cos( 4 \sqrt{K} \phi)$
for NN dimerization and of $sin ( 4 \sqrt{K} \phi )$ for NNN dimerization because
the dimerization strength is larger than the other couplings of the system,
Hence it produces a deeper minima for the system.
This parameter dependent transition, from massless phase to massive phase, 
at $T=0$ is
the quantum phase transition. 
This quantum phase transition occurs at the 
every magnetization state. So we conclude that the appearance of quantum spin
pumping is not spontaneous like $m=0$, rather dependent on the strength of the
parameter.  

\subsection{ 2.3 Calculations and Results for $m=\frac{1}{2}$ 
and Others Fractionally Quantized Magnetization States:}

Now we discuss the saturation magnetization at $m= \frac{1}{2}$ ($k_F ~=~0$).  
$K_F ~=~0$ implies that the band is empty and the dispersion is not linear, so the
validity of the continuum field theory is questionable. Values of the two 
Luttinger
liquid parameters, $v_0$ and $K$, are $0$ and $1$ respectively. 
It also implies that none of the sine-Gordon coupling terms become
relevant in this parameter
space. So there is no spin pumping for these fractionally 
quantized magnetization states.\\

Here we present the explanation for the absence of other fractionally 
quantized magnetization state (like $1/3, 1/5, 1/7$ etc):
A careful examination of Eq. 12 to Eq. 14 reveals that to get a non oscillatory
contribution from Hamiltonian one has to be satisfied $4 {k_F} = G$ condition
but this condition is not fulfilled for these fractionally quantized magnetization
state. There are no sine-Gordon coupling terms. Hence there is no spin pumping
physics for these fractionally quantized states of magnetization. 

\section{ 3. Conclusions:}
We have presented the physics of quantum spin pump for different magnetization
state of an antiferomagnetic spin-1/2 chain with competing 
exchange interactions along with bond-alternation interactions. 
Our study is completely analytical. 
We have observed that for some magnetization
state spin-pumping is spontaneous and for some other it is not and
also explain the physical reasons for the presence and absence of
spin pumping for those states.
\\  

The author (SS) would like to acknowledge The Center for Condensed Matter 
Theory of IISc for
providing the working space and The National Center for Theoretical Science 
(Taipei) where
the initial phase of this work has started. Finally author thanks, 
Dr. B. Mukhopadhyay for reading the manuscript very critically.


\begin{references}
\bibitem{thou1} D. J. Thouless, Phys. Rev. B {\bf 27}, 6083 (1983).

\bibitem{thou2} Q. Niu. Q and  D. J. Thouless, J. Phys. A {\bf 17}, 2453 (1984).

\bibitem{brou} P. W. Brouwer, Phys. Rev. B {\bf 58}, 10135 (1998).

\bibitem{alt} T. A. Shutenko, I. L. Aleiner and B. L. Altshuler, 
Phys. Rev. B {\bf 61}, 10366 (2000).

\bibitem{dot} Y. Levinson, Entin-O.Wohlman and P. Wolfe, Physica A
{\bf 302}, 335 (2001); Wohlman-O.Entin  and A. Aharony, Phys. Rev. B
{\bf 66}, 35329 (2002).

\bibitem{bla} M. Blaauboer, Phys. Rev. B {\bf 65}, 235318 (2002).

\bibitem{wire} J. Wang and B. Wang, Phys. Rev. B {\bf 65}, 153311 (2002);
B. Wang and J. Wang, Phys. Rev. B {\bf 66}, 201305 (2002).

\bibitem{lutt1} P. Sharma and C. Chamon, Phys. Rev. Lett. {\bf 87}, 96401 (2001);
P. Sharma and C. Chamon cond-mat/0209291.

\bibitem{lutt2} R. Citro, N. Anderi and Q. Niu, cond-mat/0306181.

\bibitem{shin} R. Shindou, J. Phys. Soc. Jpn {\bf 74}, 1214 (2005).

\bibitem{koh} M. Kohgi $et~al.$, Phys. Rev. Lett {\bf 86}, 2439 (2001).

\bibitem{ful} P. Fulde, B. Schmidt and P. Thalmeier, Europhys. Lett
{\bf 31}, 323 (1995).

\bibitem{shin2} Here the system is invariant under the $\pi$-rotational
symmetry which exchange the NN bonds. The direction of electric field and
$\pi$ rotational axis are different.
 
\bibitem{hal} F. D. M. Haldane, Phys. Rev. B {\bf 25}, 4925 (1982).

\bibitem{aff} S. R. White and I. Afflek, Phys. Rev. B {\bf 54}, 9862 (1996).

\bibitem{tot} K. Totsuka, Phys. Rev. B {\bf 57}, 3454 (1998).

\bibitem{tone} T. Tonegawa $et~al.$ Physica B {\bf 246-247}, 368 (1998).

\bibitem{chen} S. Chen, H. Buttner and J. Voit, Phys. Rev. Lett {\bf 87},
087205 (2001).

\bibitem{cap} L. Capriotti $et~al.$ Phys. Rev. Lett {\bf 89}, 149701 (2002).

\bibitem{sar1} S. Sarkar and D. Sen, Phys. Rev. B {\bf 65}, 172408 (2002).

\bibitem{cal} $ {S_n}^{z}~=~ \frac{1}{2 \pi} {{\partial}_x}{\phi ({x_n})}
- \frac{{(-1)}^n}{\pi \alpha} cos (\phi (x_n) )$. $\phi$ field corresponds
to the quantum fluctuations (boson) of spin.  

\bibitem{berry} M. V. Berry, Proc. R. Soc. Lond. A {\bf 392}, 45 (1984).

\bibitem{avron} J. E. Avron, A. Raveh and  B. Zur, Rev. Mod. Phys. {\bf 60}, 
873 (1988);
J. E. Avron, J. Berger and Y. Last, Phys. Rev. Lett. {\bf 78}, 511 (1997).
  
\bibitem{hrk} R. Chitra, S. Pati, H. R. Krishnamurthy, D. Sen and
S. Ramasesha, Phys. Rev. B {\bf 52}, 6581 (1995).

\bibitem{gia2} Giamarchi. T, {\it Quantum Physics in One Dimension} 
(Oxford Science Publications,
Clarendon Press, Oxford, 2004).

\bibitem{nom} K. Nomura and K. Okamoto, J. Phys. Soc. Jpn. {\bf 62}, 1123 (1993).

\bibitem{ali} R. D. Somma and A. A. Aligia, Phys. Rev. B {\bf 64}, 24410 (2001).

\bibitem{kos} J. M. Kosterlitz, and  D. J. Thouless, J. Phys. C {\bf 6}, 
1181 (1973);
V. L. Berezinski, Sov. Phys. JETP {\bf 32}, 493 (1971).

\end{references}
\end{document}